\documentstyle[osa,twocolumn,graphicx]{revtex}
\newcommand{\VV}[1]{\mbox{\boldmath{${#1}$}}}

\newcommand{\Dif}[2]{\frac{\mathrm{d}#1}{\mathrm{d}#2}}
\newcommand{\nab}[0]{\VV{\nabla}}
\begin{document}
\title{Particle Acceleration in three dimensional Reconnection Regions: A New
Test Particle Approach}
\author{R\"udiger Schopper}
\address{Geophysical Institute, University of Alaska Fairbanks,
        Fairbanks AK 99775, USA}
\author{Guido T.~Birk}
\address{Department of Astronomy, University of Toronto,
        Toronto, Ontario M5S 1A7, Canada}
\author{Harald Lesch}
\address{Institut f\"ur Astronomie und Astrophysik, LMU-M\"unchen,
        D-81679 M\"unchen, Germany}
\maketitle
\begin{abstract}
Magnetic Reconnection is an efficient and fast acceleration mechanism by means
of direct electric field acceleration parallel to the magnetic field.  Thus,
acceleration of particles in reconnection regions is a very important topic in
plasma astrophysics.  This paper shows that the conventional analytical models
and numerical test particle investigations can be misleading concerning the
energy distribution of the accelerated particles, since they oversimplify the
electric field structure by the assumption that the field is homogeneous.  These
investigations of the acceleration of charged test particles are extended by
considering three-dimensional field configurations characterized by localized
field-aligned electric fields.  Moreover, effects of radiative losses are
discussed.  The comparison between homogeneous and inhomogeneous electric field
acceleration in reconnection regions shows dramatic differences concerning both,
the maximum particle energy and the form of the energy distribution.
\end{abstract}

\noindent\rule{\linewidth}{0.1pt}

\section{Introduction}\label{sec:Intro}

The acceleration of charged particles in space and astrophysical plasmas is a
matter of debate in contemporary plasma astrophysics (cf.~reviews~\cite{Kirk94}
and~\cite{Kuijpers96}).  Beside the "standard" acceleration mechanisms as shock
waves (\cite{Ginzburg64}, \cite{Blandford78}), Fermi
acceleration~\cite{Fermi49}, and magnetic turbulence~(\cite{Ramaty87},
\cite{Schlickeiser89}) acceleration by magnetic field-aligned electric fields
receives more and more attention as it offers the capability for efficient and
fast energization of relativistic particles~(\cite{Alfven81},
\cite{Schindler91}).  The formation and amplification of such electric fields in
the context of magnetic reconnection processes has been discussed for different
cosmical plasmas and parameter regimes~( \cite{Priest87}, \cite{Schindler91},
\cite{Otto93}, \cite{Lesch97}, \cite{Kopp97}).  Since magnetic reconnection can
be regarded as an unavoidable dissipation and relaxation process in magnetized
plasmas under the influence of externally driven shear motions, the principal
importance of magnetic reconnection electric fields is now widely accepted.

Analytical and numerical studies have significantly improved our understanding
of energetic particle trajectories in reconnection configurations
without~(\cite{Speiser65}, \cite{Cowley78}, \cite{Deeg91}, \cite{Moses93},
\cite{Vekstein95}, \cite{Kliem98}) and with~(\cite{Litvinenko93},
\cite{Kaufmann94}, \cite{Litvinenko96}) a magnetic field component parallel to
the reconnection current sheet.  Nevertheless, the particle trajectories and, in
particular, the actual energy gains of accelerated charged particles in the
reconnection current sheets prove to be still a crucial matter of debate, since
the studies mentioned above have been performed for idealized analytical
electric and magnetic field configurations which differ significantly from the
three-dimensional configurations characterized by localized field-aligned
electric fields as observed, e.g.~in the solar corona (cf.~\cite{Bentley97}) and
the Earth's magnetosphere (cf.~\cite{Akasofu81}). Test particle simulations of
ion (\cite{Birn97}) and electron (\cite{Birn98}) injection in a high-beta
magnetospheric substorm configurations have shown that the consideration of more
complicated 3d field configurations rather than idealized 2D ones is crucial for
the understanding of particle acceleration in reconnection regions.  In this
contribution we study the acceleration of charged particles up to relativistic
energies in reconnection regions.  The electromagnetic fields of the
reconnection region are modeled by 3D-magnetohydrodynamic (MHD) simulations of a
reconnection zone in a current carrying magnetic flux tube.  Since the
astrophysical evidence for the existence of relativistic particles relies only
on the electromagnetic radiation emitted by of these particles, we calculate the
motion of high-energy particles by means of relativistic test particle
simulations including radiative forces due to inverse Compton scattering and
synchrotron radiation.  We consider our study as a step forward to a
self-consistent description of particle acceleration in quite realistic
reconnection configurations.

The manuscript is organized as follows: The next section presents the motivation
for our new approach.  Sec.~\ref{sec:MHD} describes the electromagnetic field
configuration we used and how they are obtained from MHD simulations.  These
fields represent the external force field, to be supplemented by radiative
forces, for the relativistic test particle calculations, described in
Sec.~\ref{sec:Accel}.  The results of these test particle simulations are
compared with the analytical and numerical X-type configuration of the
reconnection magnetic field in Sec.~\ref{sec:Comp}.  Sec.~\ref{sec:Disc} is
devoted to a discussion of our results.

\section{Motivation for our new Approach}\label{sec:Review}

Particle acceleration in magnetic reconnection zones has been extensively
discussed and studied analytically and numerically (we refer to the analytical
papers \cite{Speiser65}, \cite{Bulanov75}, \cite{Bulanov80},
\cite{Litvinenko96}, and references therein); numerical treatments can be found
in \cite{Deeg91}, \cite{Nocera96}, and references therein).  These
investigations rely on one common assumption The homogeneity of the electric
field.  The authors have used an analytical formulation of the electric field
configuration and therefore the calculations are highly idealized.  The field
configurations found in the papers mentioned above are nearly identical and can
be written as
\begin{equation}
  \VV{B}\left( \VV{r} \right) = B_0
  \left(
  \begin{array}{cc}%
     \frac{y}{L_x}                  \\
     \frac{x}{L_y} & \!\!\!\!\!+\,\xi \\
     1
  \end{array}
  \right)\quad
  \VV{E}\left( \VV{r} \right) = E_0
  \left(
  \begin{array}{c}%
     0 \\
     0 \\
     1
  \end{array}
  \right).
\end{equation}
The process of reconnection in two dimensions is still subject of intense
scientific investigation up to the present day.  Besides first analytical
approaches (\cite{Demoulin96}; \cite{Hornig98}) investigations of the much more
complex 3D reconnection process are restricted to numerical modeling.  In a 2D
description the X--point configuration
\begin{equation}
  \VV{B} \left( \VV{r} \right) = B_0
  \left(
  \begin{array}{c}%
     \frac{y}{L_x} \\
     \frac{x}{L_y} \\
     0
  \end{array}
  \right), \label{equ:xfield}
\end{equation}
that describes a hyperbolic magnetic field with a neutral line in the center
($\VV{B}(0,0,z) = 0$), has become the standard model of reconnection.  With
Ampère's Law
\begin{equation}
  \frac{4\pi}{c}\VV{j} = \nab \times \VV{B},
\end{equation}
Ohm's Law
\begin{equation}
  \eta \VV{j} = \VV{E} + \VV{v}/c \times \VV{B},
\end{equation}
and invariance in one dimension, say $\partial_z\cdot = 0$, one can easily
derive the electric field
\begin{equation}
   \VV{E} = \eta \frac{B_0 c}{4\pi} \left( \frac{1}{L_x} -
   \frac{1}{L_y} \right)
   \left(
   \begin{array}{c}%
      0 \\
      0 \\
      1 \\
   \end{array}
   \right)
   -\VV{v}/c \times \VV{B}.  \label{equ:efield}
\end{equation}
If one assumes a constant resistivity and that the convective term $\VV{v}/c
\times \VV{B}$~is small as compared to the first term in~(Equ.~\ref{equ:efield})
in the localized reconnection zone the constant, homogeneous, infinitely
extended electric field is given by
\begin{equation}
   \VV{E} = \eta \frac{B_0 c}{4\pi} \left( \frac{1}{L_x} -
   \frac{1}{L_y} \right)
   \left(
   \begin{array}{c}%
      0 \\
      0 \\
      1 \\
   \end{array}
   \right).
\end{equation}
Such an electric field is, without question, unphysical.  In case of a 2D
reconnection calculation this is not a severe problem, because the electric
field is never used as a real physical quantity.  For this reason one can argue
that it is merely a mathematical auxiliary variable and only the magnetic field
is needed to describe the complete electromagnetic configuration.

However, this X--type field is a valid approximation of the inner part of a
reconnection region around the neutral line.  Only the application of this
idealized configuration in the context of particle acceleration gives rise to
serious problems.  The electric field is the source of energy for the
accelerated particles and exactly this very reason for acceleration is, as
stated above, unphysical.  At this point, we should ask how sensible are the
results concerning particle energy spectra, duration of acceleration etc.,
derived from an idealized, homogeneous electrical field? For example, there are
particle trajectories
\begin{equation}
   \VV{r} = \left(\begin{array}{c}%
                     0 \\
                     0 \\
                     z
                  \end{array}%
            \right)\quad%
   \VV{p} = \left(\begin{array}{c}%
                     0 \\
                     0 \\
                     p
                  \end{array}%
            \right)\quad%
   \VV{B} = B_0 \left(\begin{array}{c}%
                         \frac{y}{L_x} \\
                         \frac{x}{L_y} \\
                         1
                      \end{array}%
                \right)
\end{equation}
for which particles will gain infinite energy.  This comes from the fact that
the electric field lacks any dependence on the invariant direction.

In order to accelerate particles efficiently a strong guiding
component~$B_z$~(\cite{Bulanov75}) is needed.  With such a strong~$B_z$
component the magnetic field is almost parallel to the electric field in a
region $B_z^2 \gg B_x^2 + B_y^2$ around the neutral line.  In addition, this
guiding component captures the particles in the zone, at least as long as their
gyroradius~$r_{\rm ce}$ is smaller than the width of the reconnection zone.
These two facts guarantee the efficient acceleration described
in~\cite{Bulanov75}.  On the other hand this configuration in the center of a
X~field with strong~$B_z$ resembles remarkably that of a simple capacitor.  Any
particle following the field lines is accelerated exactly linear, like in a
capacitor.  The energy gained by such a particle is directly given by
\begin{equation}
   \Delta{\cal E} = q E_z \Delta L.
\end{equation}
$\Delta{\cal E}$ denotes the energy gain and $q$~is the charge of the particle
that is accelerated along the magnetic field line over the distance $\Delta L$.
  \begin{figure}[t]
     \center
     \includegraphics[width=0.8\linewidth,keepaspectratio]{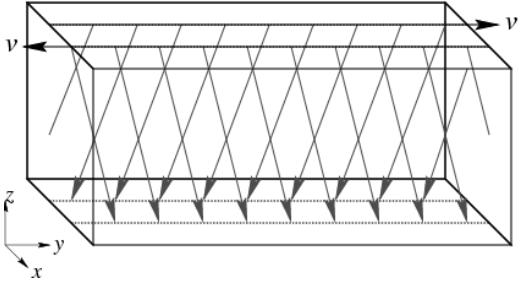}
     \caption{Sketch of the generic initial configuration.  }
     \label{fig:initial}
  \end{figure}
The energy gain does not depend on the detail of the reconnection process, but
on the geometry of the acceleration region only.

Moreover, one has to take into account the convective electric field
components~$E_x$ and $E_y$, which are completely neglected by most authors,
though they naturally arise in realistic descriptions of reconnection zones.
Such fields are generated by external shear flows which are the macroscopic
generator of electromagnetic energy converted partly into particle kinetic
energy in the localized dissipative reconnection zones.  Calculations with
three-dimensional magnetohydrodynamic~(3D-MHD) codes clearly show
(e.g.~\cite{Birk97a}) that those components are often much stronger than the
field line parallel electric field~$E_\parallel$.  The precise influence of
those components is not known, nevertheless we should not neglect them strait
from the beginning, without knowing anything about the strength of their
influence.  We speculate that such convective electric fields can filter out
particles from an efficient acceleration due to their strength and structure and
dependent on the initial positions of the particles.  In other words, the form
of the energy distribution may be closely connected to these electric fields.

The above questions convinced us to approach again this well known and well
studied phenomenon of particle acceleration in magnetic reconnection zones.
Contrary to other investigations, we use a field configuration, which is
produced with the help of a full 3D-MHD code.

\section{The 3D-MHD Simulations of Magnetic Reconnection}\label{sec:MHD}
The MHD framework for the particle simulations of electron acceleration in
reconnection regions is built up by means of a three-dimensional MHD
code~\cite{Otto90}.  It is motivated by previous works on reconnection in the
context of planetary magnetosphere-ionosphere coupling~(\cite{Otto93},
\cite{Kopp97}, \cite{Litvinenko96}, \cite{Birk97a}, \cite{Birk97b}) and of
active galactic nuclei~\cite{Lesch97}.  However, the situation we have in mind
(cf.~Fig.~\ref{fig:initial}) is quite general and may be regarded as a rather
generic reconnection configuration that arises when MHD shear flows working as
electromagnetic generators sheared magnetic fields.  For convenience, we start
from a two-dimensional force-free magnetic field
\begin{equation}
   \VV{B} = B_{y0}\tanh(x)\;\VV{\rm e}_y\ -
             \sqrt{{B_{z0}}^2 + {{B_{y0}}^2\over\cosh^2(x)}}\;\VV{\rm e}_z.
\end{equation}
We choose for the constant main component and the shear component of the
magnetic field $B_{z0}=5$ and $B_{y0}=1$ in normalized units.  The plasma is
chosen to be isothermal as well as homogeneous, initially.  We note that,
alternatively, an appropriate equilibrium that includes thermal pressure forces
could be realized.  This ideal equilibrium is perturbed by
\begin{equation}
  v_y(x,z) = v_{y0}{\tanh(2x)\over\cosh^2\left({x\over 3}\right)}
               \exp\left(-{z \over 6}\right).
\end{equation}
As a boundary condition for $t>0$ we use
\begin{equation}
   v_y(x,z=z_{\rm min})=v_{y0}{\tanh(2x) \over \cosh^2\left({x
   \over 3}\right)}.
\end{equation}
These velocity perturbation with an amplitude chosen as $v_{y0} = 0.5 \%$ of the
Alfv\'en velocity mimics some external convective plasma motion which leads to a
further shear of the magnetic field and an increase of the field-aligned current
density.  The associated magnetic perturbation is transported along the
$z$-component of magnetic field via shear Alfv\'en waves.  Here, we follow the
general concept that plasma systems that are compelled by such external forces
when some critical parameter(s) is (are) exceeded non-linearly develop into a
state of lower energy thereby releasing stored free energy in form of
dissipative plasma heating and particle acceleration.  
  \begin{figure}[bp]
     \center
     \includegraphics[angle=90,width=0.8\linewidth,keepaspectratio]{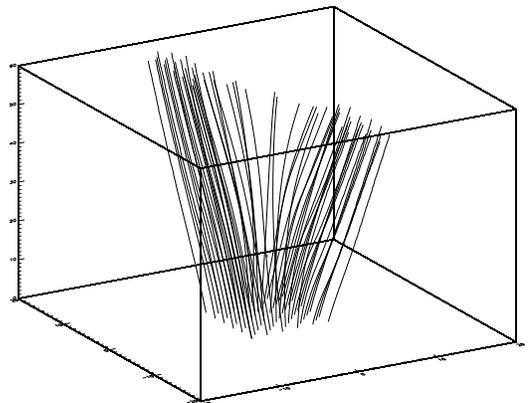}
     \caption{Some characteristic magnetic fieldlines after~$t=120\tau_{\rm A}$  }
     \label{fig:fieldlines}
  \end{figure}
  \begin{figure}[tp]
     \center
     \includegraphics[width=0.8\linewidth,keepaspectratio]{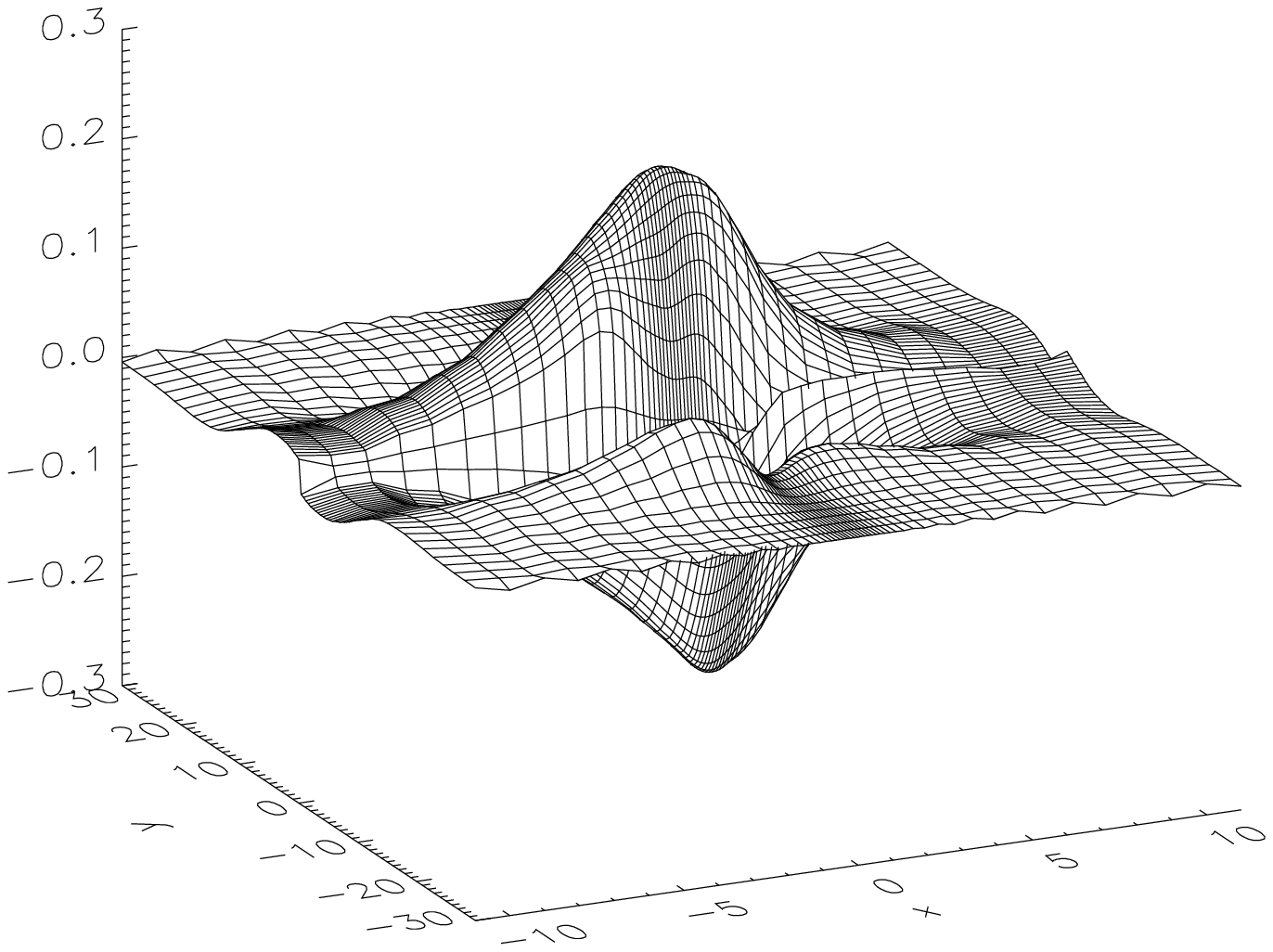}
     \includegraphics[width=0.8\linewidth,keepaspectratio]{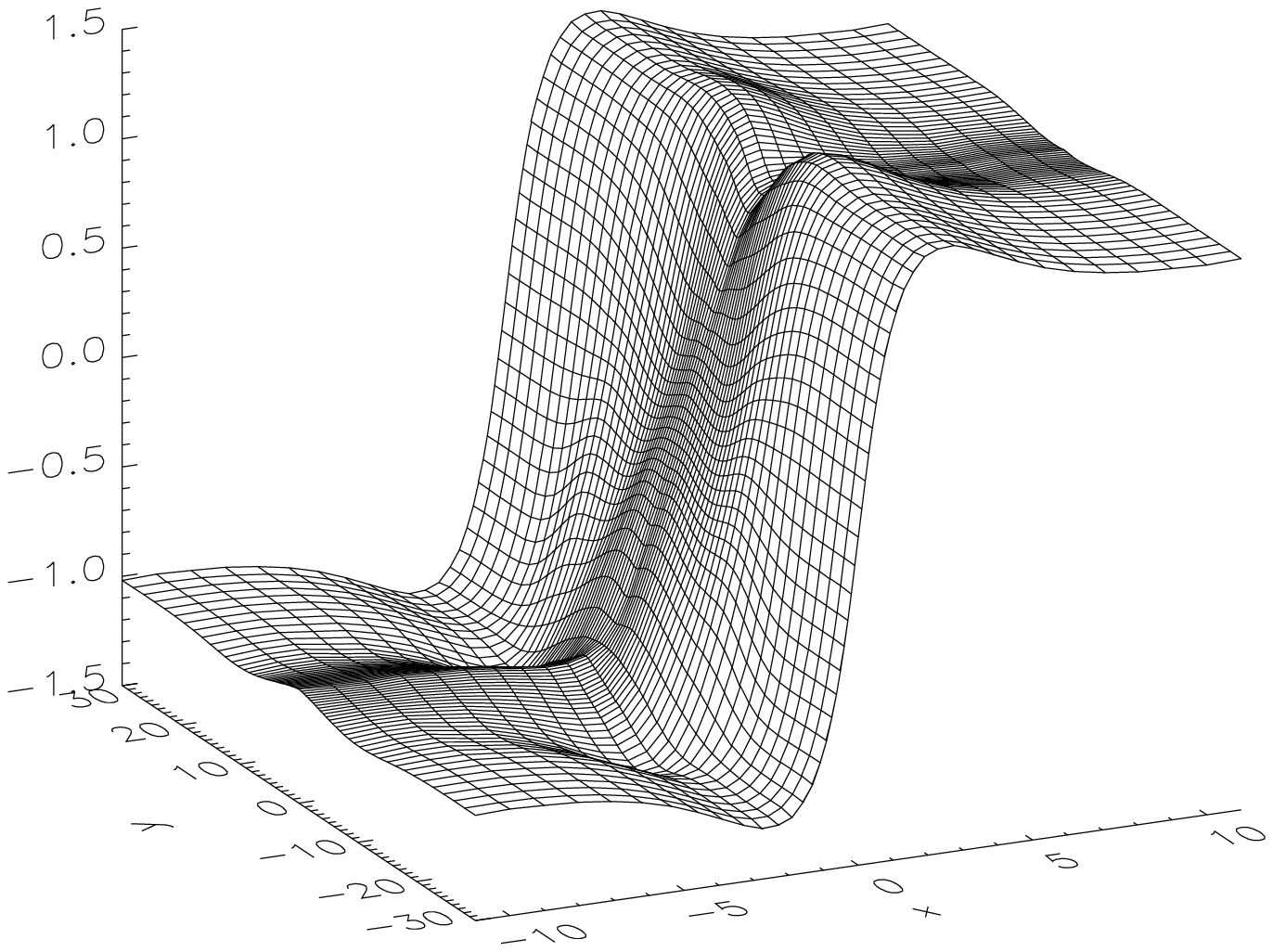}
     \includegraphics[width=0.8\linewidth,keepaspectratio]{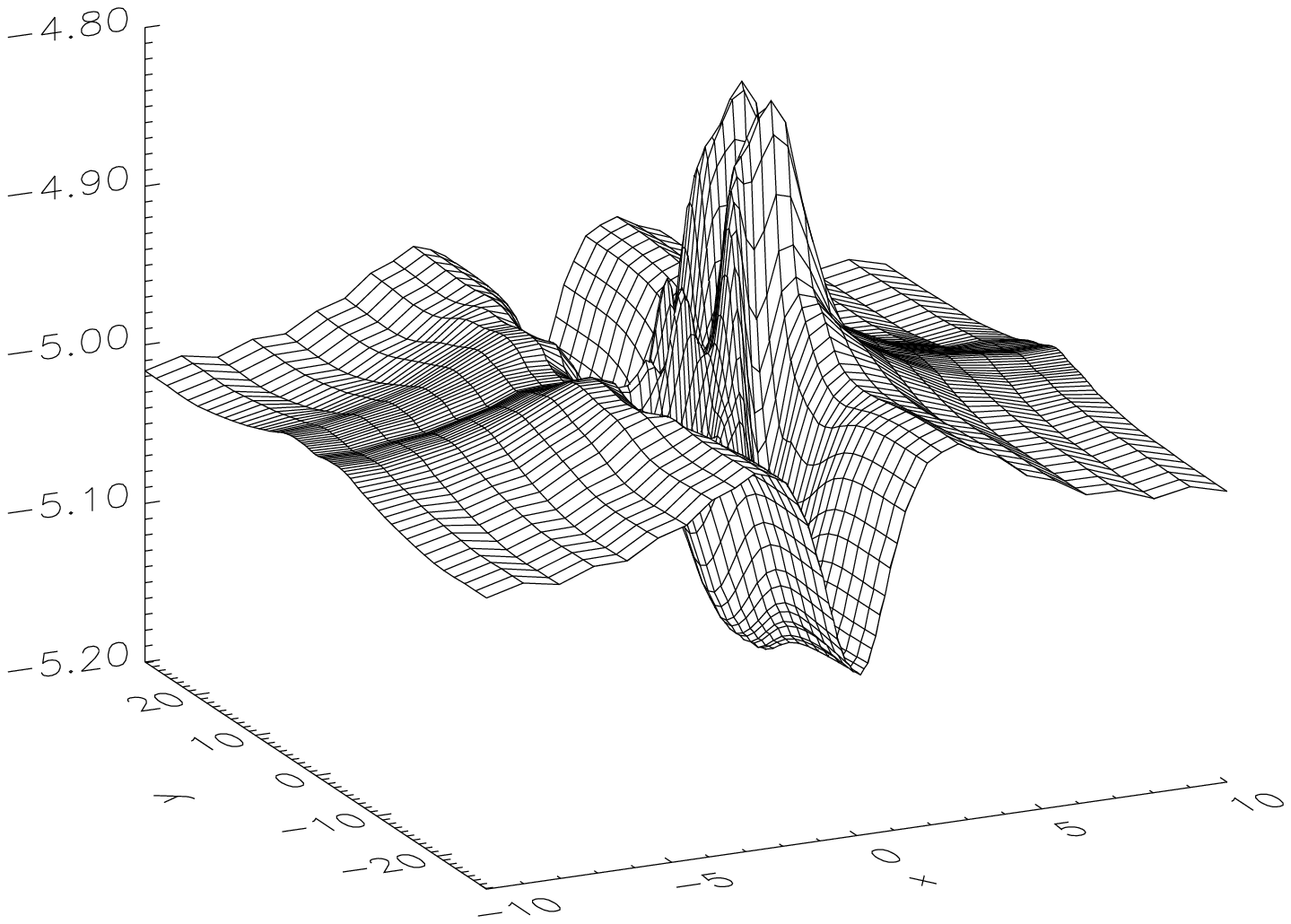}
     \caption{Snapshot of the $x$-, $y$- and $z$-component (from top to
              bottom) of the reconnection magnetic field at the height of
              the central reconnection region after~$t=120\tau_{\rm A}$.  }
     \label{fig:bxbybza}
  \end{figure}
In the considered configuration the magnetic field-aligned current density plays
the role of the critical parameter.  When the current density exceeds a critical
value (we choose a marginal current density at the beginning) a current
dependent resistivity, which is localized in the $y$- and $z$-direction is
switched on.  This violation of ideal Ohm's law starts the magnetic reconnection
process (cf.\cite{Schindler91}).  Fortunately, for our purpose we have not to
dwell on the specific nature of the finite resistivity, may it be
microturbulent~\cite{Huba85} dissipation or particle inertia~\cite{Vasyliunas75}
or whatever.  For the MHD simulations we choose the following technical details:
  \begin{figure}[tp]
     \center
     \includegraphics[width=0.8\linewidth,keepaspectratio]{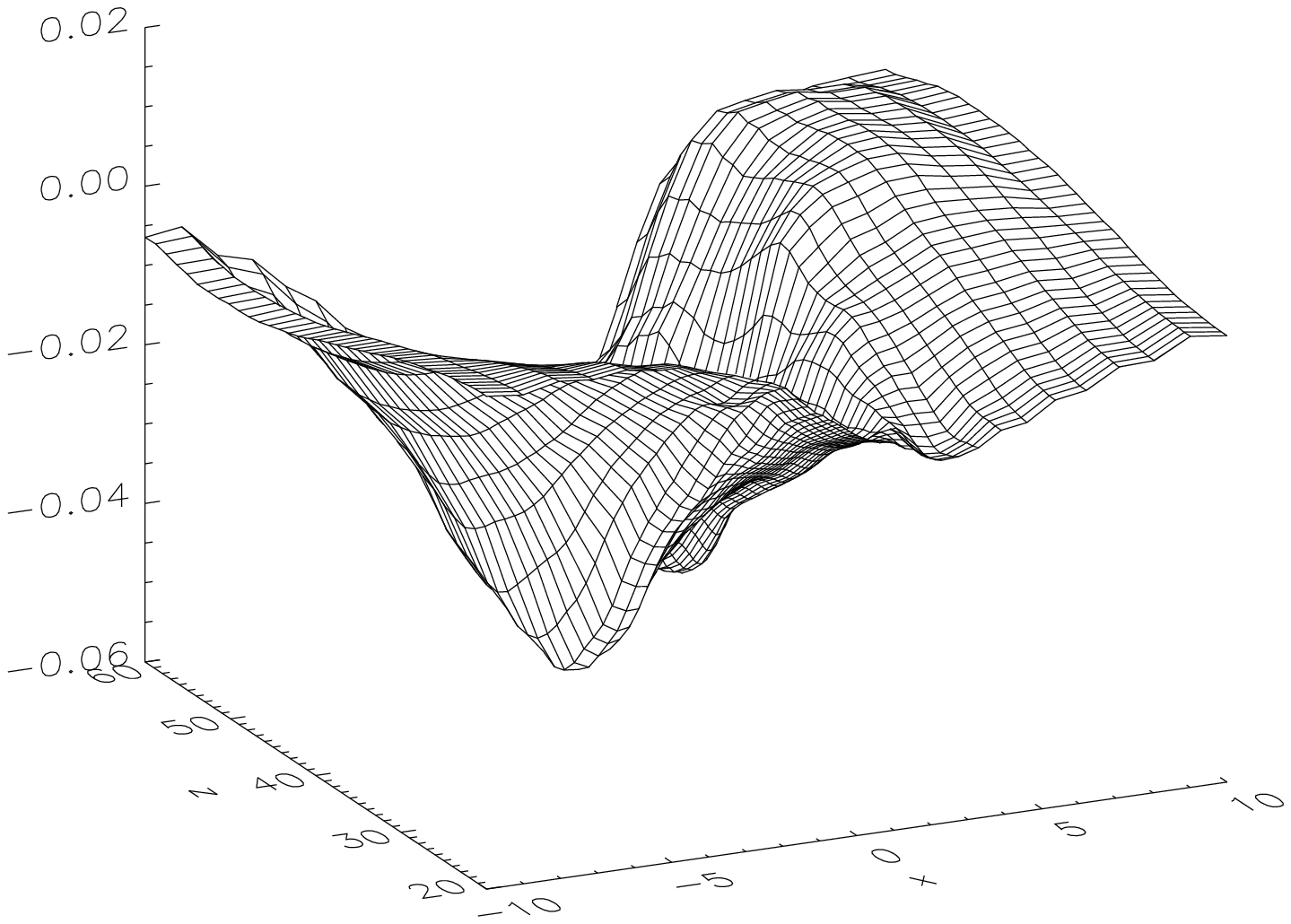}
     \includegraphics[width=0.8\linewidth,keepaspectratio]{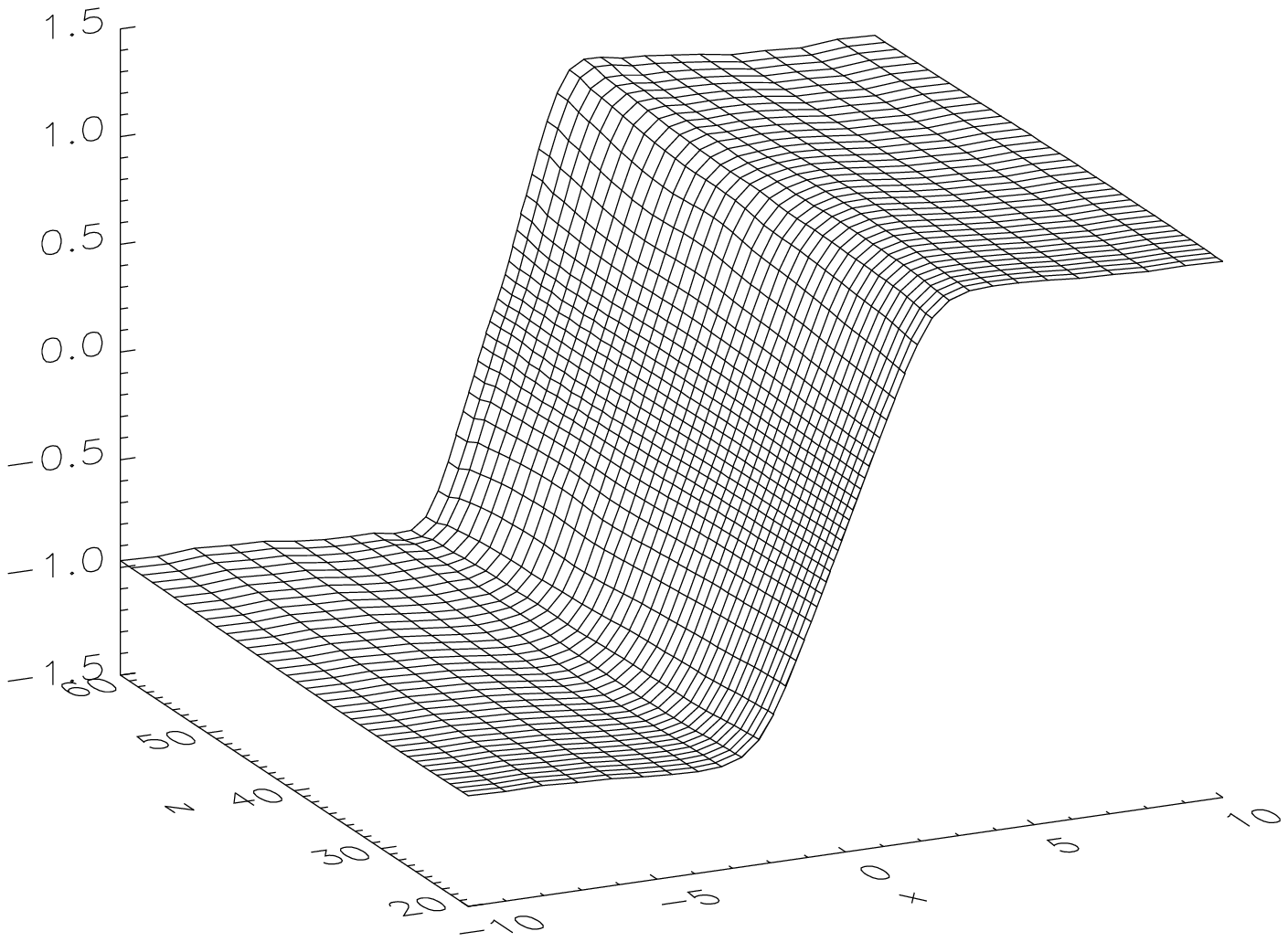}
     \includegraphics[width=0.8\linewidth,keepaspectratio]{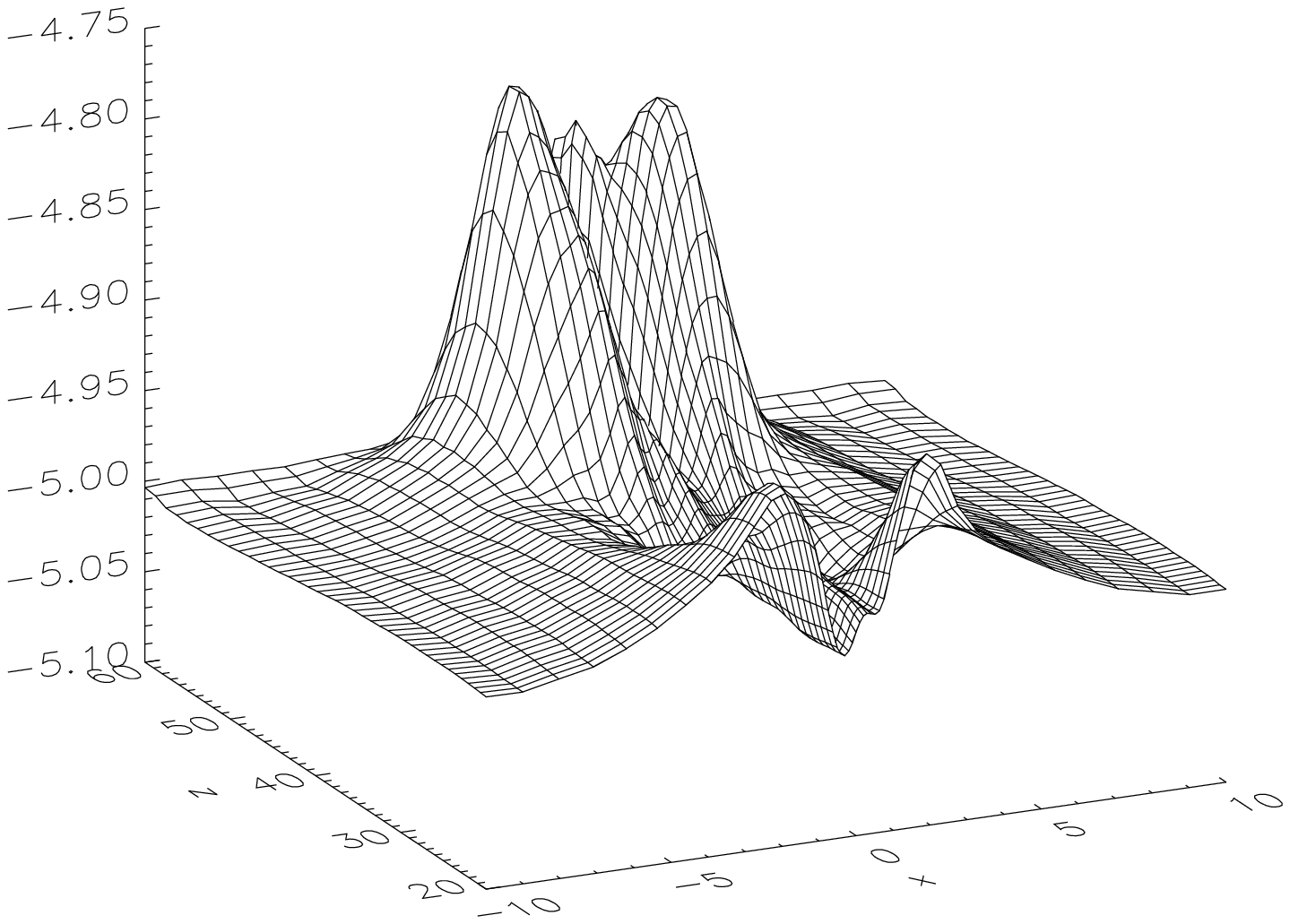}
     \caption{Snapshot of the $x$-, $y$- and $z$-component (from top to
              bottom) of the reconnection magnetic field at
              the $y=0$-half-plane after $t=120\tau_{\rm A}$.  }
     \label{fig:bxbybzb}
  \end{figure}
The dimensions of the numerical box are given by $x \in [-10,10]$, $y \in
[0,30]$, and $z \in [0,60]$ in normalized units.  We make use of line symmetry
as a boundary conditions at $y=0$, i.e.~the $x$- and $y$-components of the
magnetic field and the velocities are chosen antisymmetric whereas the
$z$-components of these quantities are chosen symmetric about $y=0$.  Boundary
conditions in $x$ are chosen as symmetric and in $z$ as well as at $y=y_{\rm
max}$ we use extrapolation.  The simulations are carried out with $49$ grid
points in the $x$- direction, $39$ grid points in the $y$-direction and $105$
grid point in the $z$-direction.  
  \begin{figure}[tp]
     \center
     \includegraphics[width=0.8\linewidth,keepaspectratio]{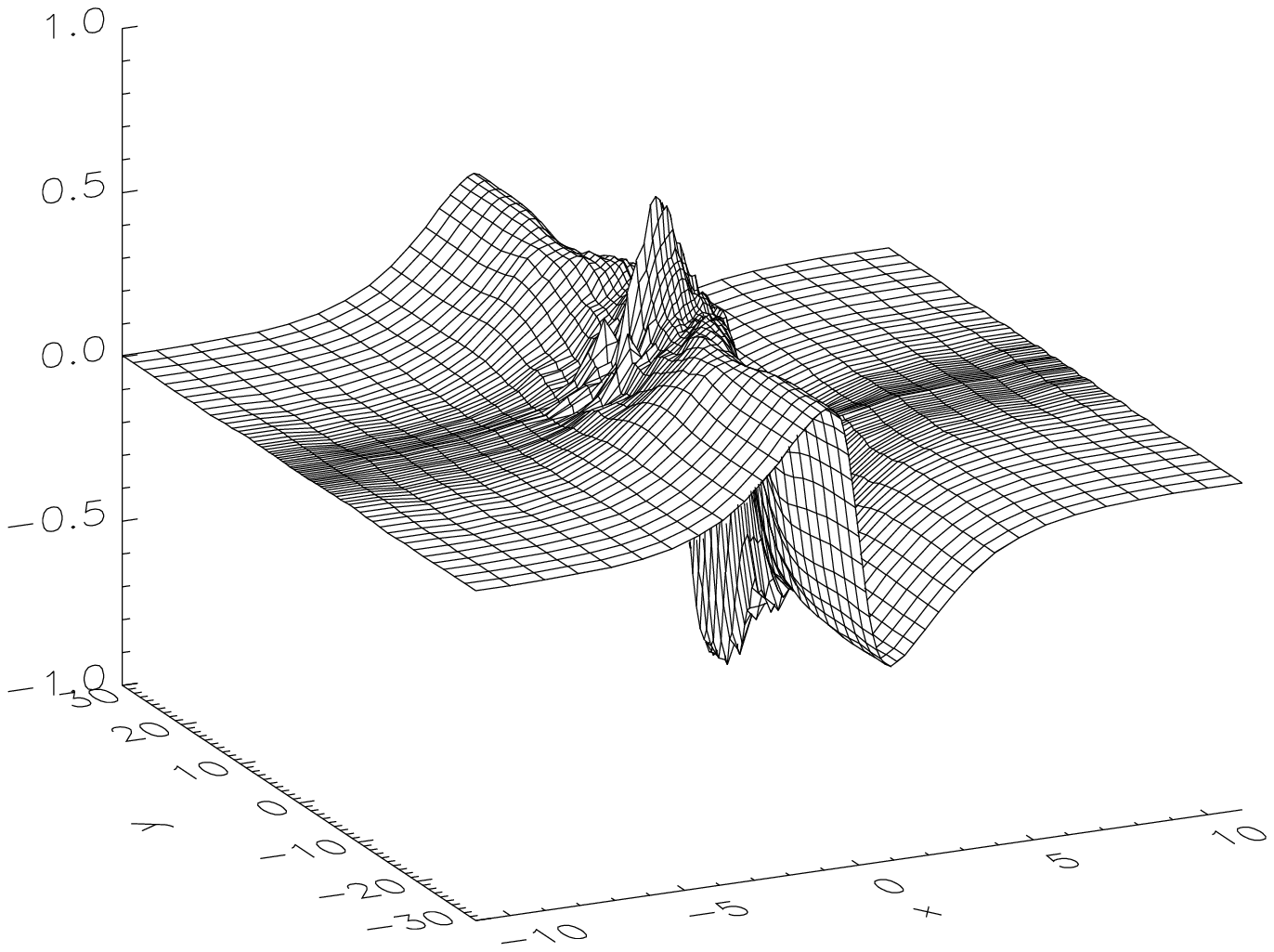}
     \includegraphics[width=0.8\linewidth,keepaspectratio]{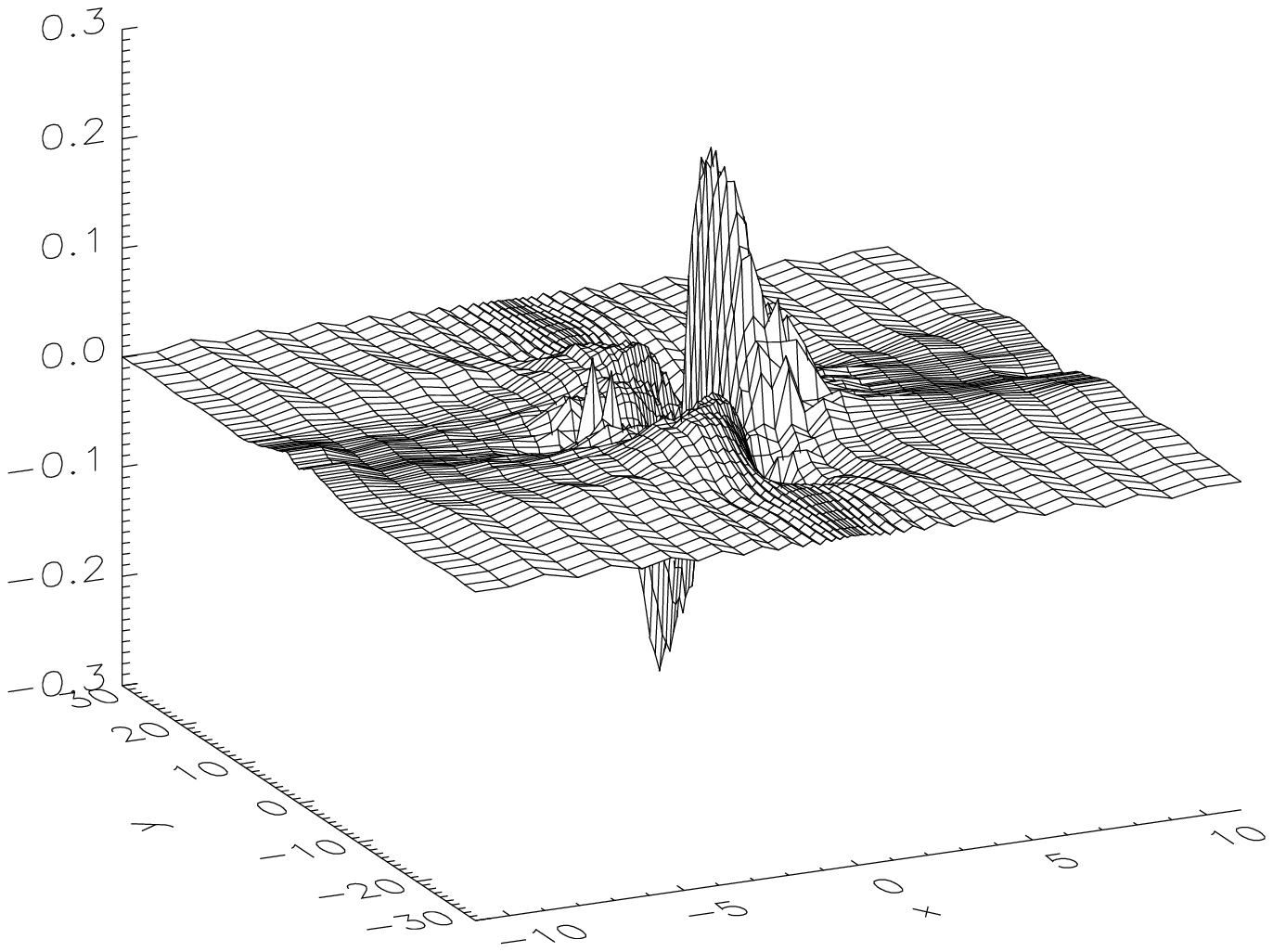}
     \includegraphics[width=0.8\linewidth,keepaspectratio]{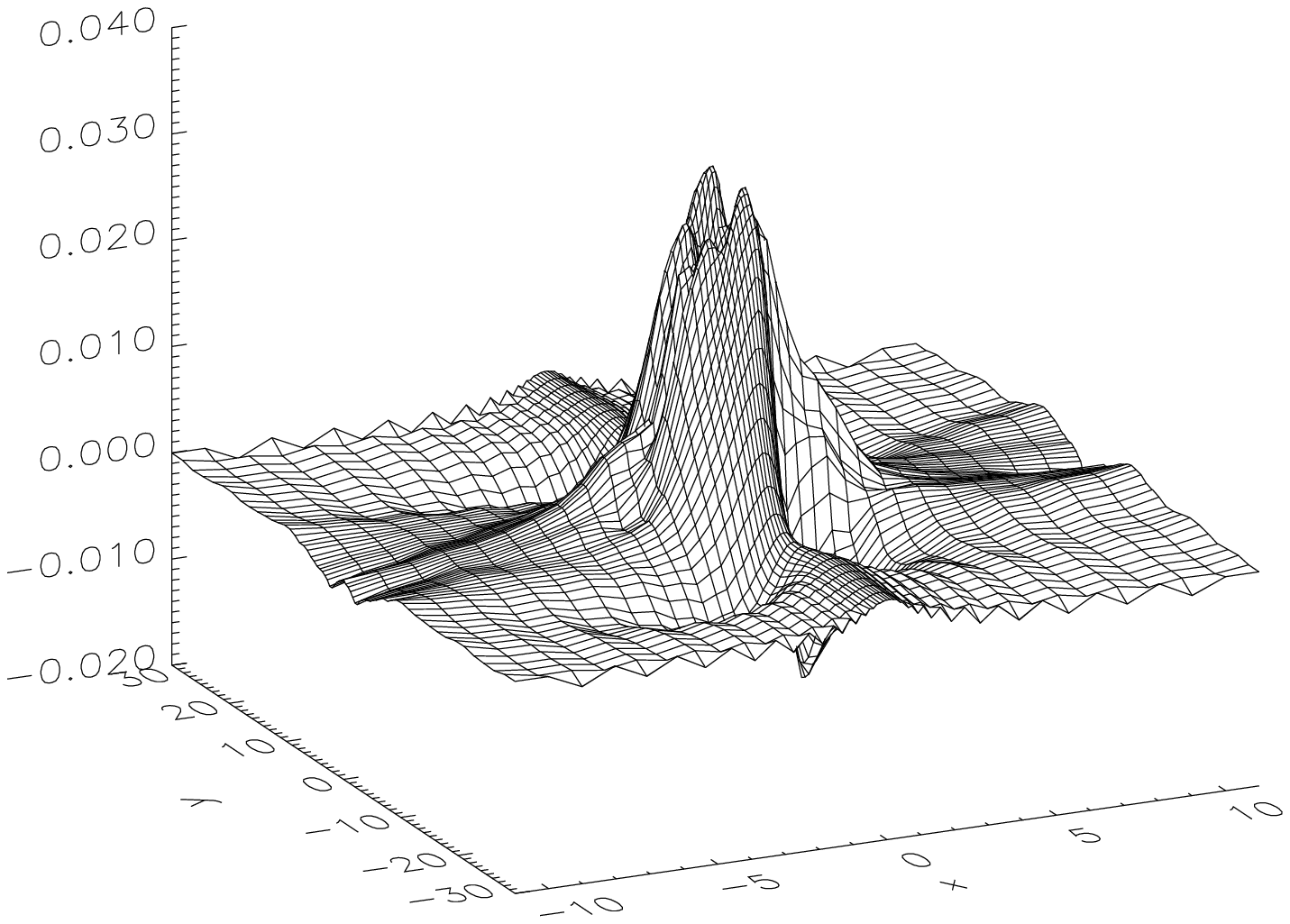}
     \caption{Snapshot of the $x$-,$y$- and $z$-component (from top to
              bottom) of the reconnection electric field at the height of
              the central reconnection region
              after~$t=120\tau_{\rm A}$.  }
     \label{fig:exeyeza}
  \end{figure}
The numerical grid is a non-uniform one with a
maximum resolution of $0.05$ in the $x$-direction, $0.4$ in the $y$-direction,
and $0.2$ in the $z$-direction.

\begin{figure}[tp]
     \center
     \includegraphics[width=0.8\linewidth,keepaspectratio]{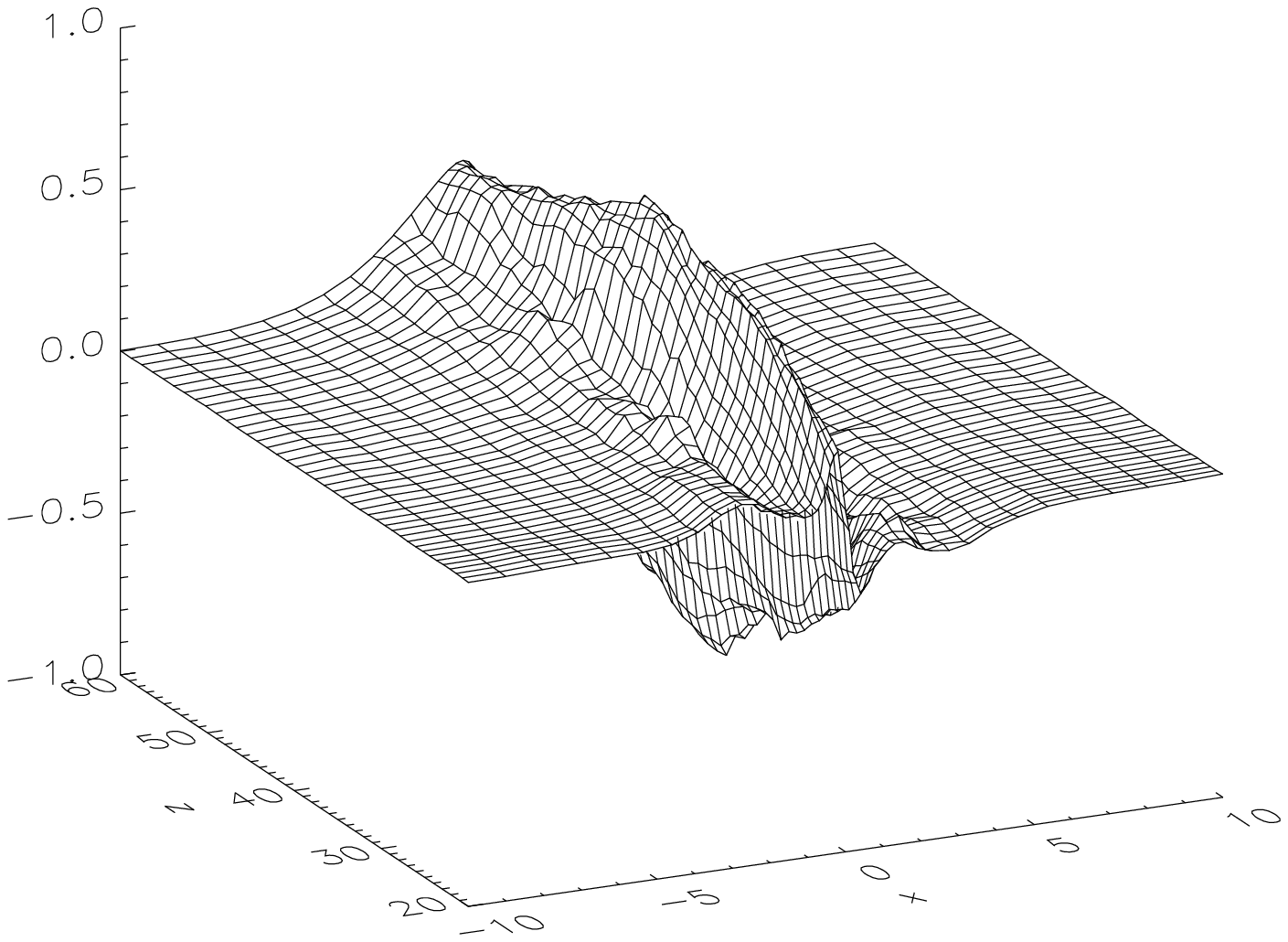}
     \includegraphics[width=0.8\linewidth,keepaspectratio]{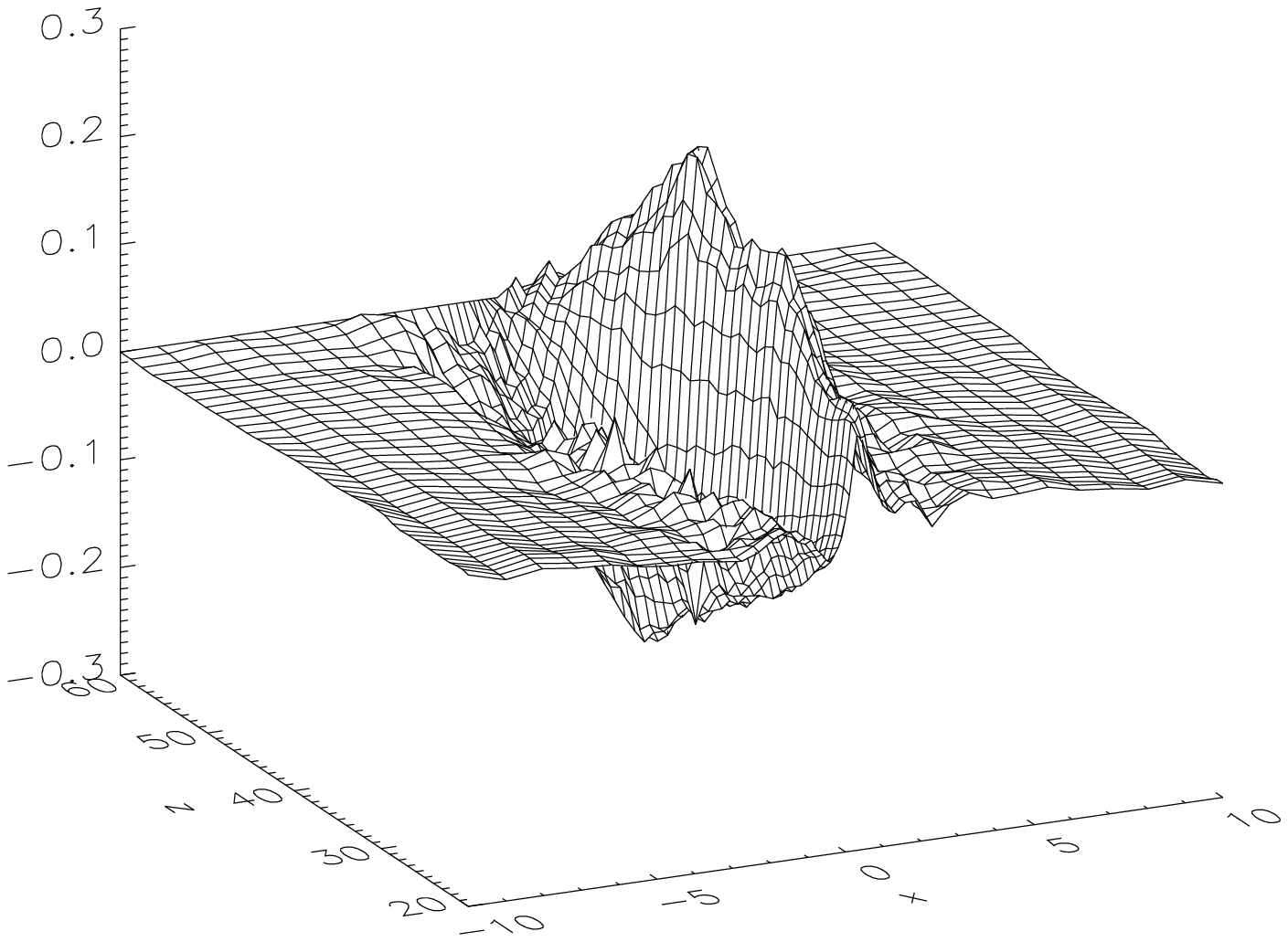}
     \includegraphics[width=0.8\linewidth,keepaspectratio]{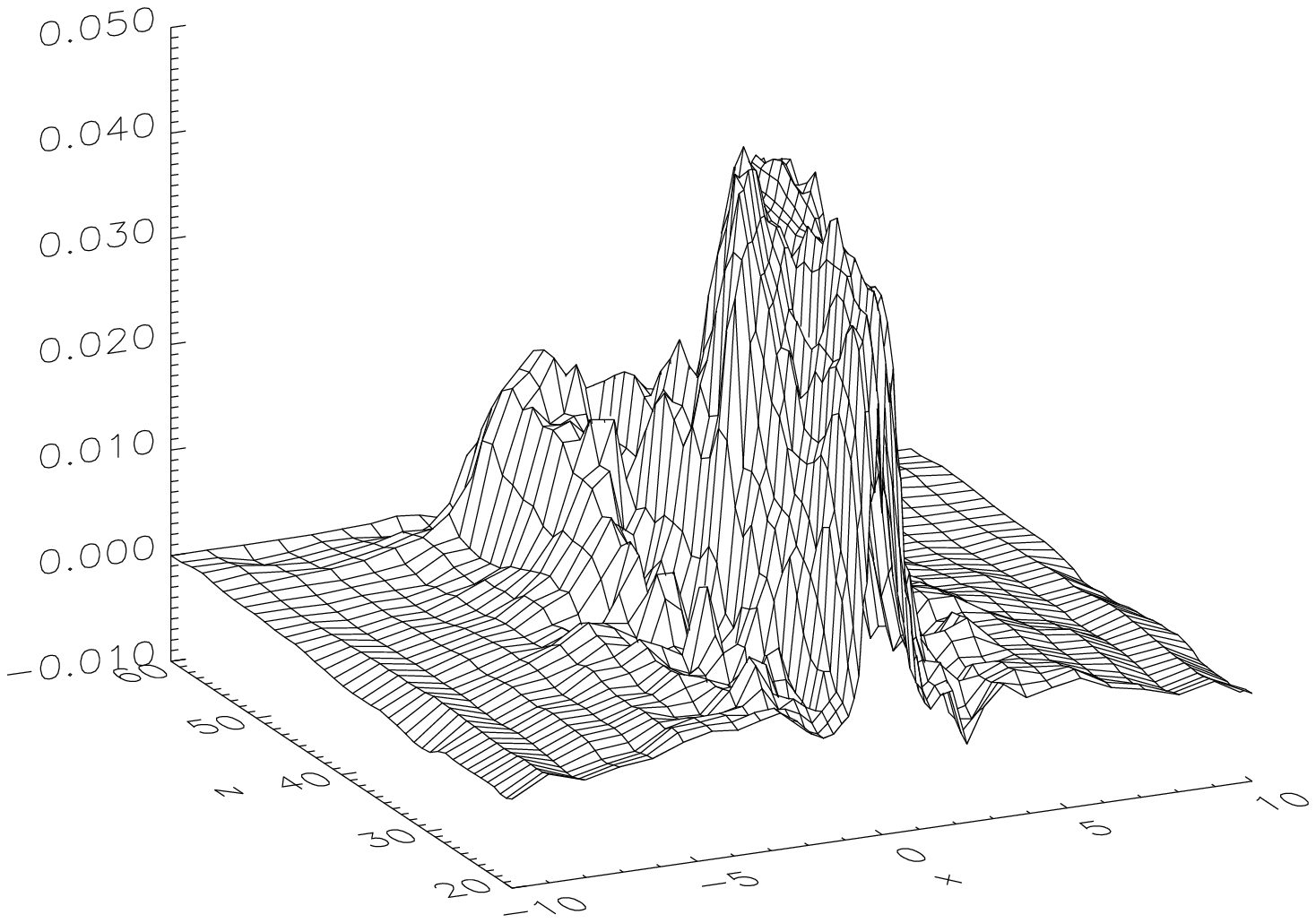}
     \caption{Snapshot of the $x$-,$y$- and $z$-component (from top to
              bottom) of the reconnection electric field at
              the $y=0$-half-plane after~$t=120\tau_{\rm A}$.  }
     \label{fig:exeyezb}
  \end{figure}

Some characteristic fieldlines are shown in Fig.~\ref{fig:fieldlines} to give an
idea of the overall structure of the magnetic field.  In Figs.~\ref{fig:bxbybza}
to \ref{fig:exeyezb} we show the Cartesian components of the electric and
magnetic fields which are chosen as an input for the particle simulation
studies.  The figures show the fields after $t=120\tau_{\rm A}$ where $\tau_{\rm
A}$ is the characteristic Alfv\'en velocity of the system.  At this state of the
temporal evolution the reconnection process can be considered as
quasi-stationary, i.e.~the magnetic flux is as rapidly piled up by the shear
flow as it is annihilated by 
magnetic reconnection.

The $x$-component of the magnetic field results from the shear flow and from the
divergent reconnection flow (Fig.~\ref{fig:bxbybza}, top panel) and
(Fig.~\ref{fig:bxbybzb}, top panel).  The $y$-component of the magnetic field
(middle panels of Fig.~\ref{fig:bxbybza} and Fig.~\ref{fig:bxbybza}) is
associated with the current sheet and modified by reconnection convergent flows
($v_x(x)$).  The $z$-component (lower panels of Fig.~\ref{fig:bxbybza} and
Fig.~\ref{fig:bxbybza}) shows a rather moderate variation due to the
reconnection process.  The three-dimensional electric vector field at the height
of the central reconnection region is illustrated in Fig.~\ref{fig:exeyeza}.
The convective $x$-component (top panels of Fig.~\ref{fig:exeyeza} and
Fig.~\ref{fig:exeyezb}) is the result of both the applied sheared flow and the
divergent reconnection flow whereas the convective $y$-component (middle panels
of Fig.~\ref{fig:exeyeza} and Fig.~\ref{fig:exeyezb}) is due to the convergent
reconnection flow.  The convective electric fields are about one order of
magnitude higher than the $z$-component of the electric field (lower panels of
Fig.~\ref{fig:exeyeza} and Fig.~\ref{fig:exeyezb}) which is caused by the finite
anomalous electric resistivity $E_z \sim \eta j_z \approx E_\parallel \approx
\eta j_\parallel$.  $E_z$ is localized in the $x$- and $y$-direction (we have
chosen for the resistivity profile $\eta \sim cosh(y(iv)/2.)$) and is most
important for the acceleration of charged particles.

 It is mainly responsible for the non-vanishing elongated generalized
field-aligned electric potential~(\cite{Schindler91}, \cite{Schindler88})
$U=-\int E_\parallel ds$ (where the integration is extended over the arch length
$s$).  Such potentials are observed, e.g.~as "auroral potential structures" in
the Earth's magnetosphere~\cite{Mozer81} which are responsible for the
acceleration of auroral electrons.  The three-dimensional electric and magnetic
vector field are used as an input for the test particle simulation studies of
relativistic electrons as discussed in the next chapter.  Note, that these
rather complex fields, which are due to shear Alfv\'en waves, shear motion and
reconnection dynamics, differ significantly from idealized configurations
usually discussed in the literature. We consider the electromagnetic field shown
as a quite typical reconnection configuration in systems that are characterized
by shear flows. Since it represents the self-consistent reconnection dynamics it
can be qualified as a realistic configuration in comparison to idealized
two-dimensional ones.

\section{Particle Acceleration}\label{sec:Accel}

We use the fields described in Sec.~\ref{sec:MHD} for electron acceleration.  By
means of a fully relativistic test particle code, including energy losses due to
synchrotron emission (SY) and inverse Compton scattering (IC) we analyze the
fate of the relativistic particles.  The term "test particle" in this context
means that any back-reaction of particles on the electromagnetic field
configuration is neglected.  As long as only a fraction of particles is actually
accelerated, i.e.~as long as the particles leave the acceleration zone on time
scales much shorter than the Alfv\'enic time this is a very good approximation,
which is always fulfilled during our simulations, since the particles are
efficiently accelerated along the magnetic field line and move with almost the
velocity of light.  Moreover, no interaction between the particles is included.

For the integration of the relativistic equations of motion
\begin{equation}
   \Dif{\VV{r}}{t} = \VV{v} \quad \Dif{\VV{p}}{t} = \VV{F} \quad
   \VV{p} = \frac{m_0\VV{v}}{\sqrt{1 - \left( \frac{v}{c} \right)^2}}
\end{equation}
we use the classical fourth order Runge-Kutta algorithm together with an
adaptive stepsize control~\cite{Press}.  Our 3D-MHD fields are only known on a
discrete 3D grid.  For this reason, we have to interpolate these fields for
positions in between by means of a simple linear interpolation algorithm.

As described in Sec.~\ref{sec:MHD} the electromagnetic fields can be regarded as
quasi-stationary after some hundred Alfvénic times~$t \simeq 120 \tau_{\rm A}$.
Moreover, the test electrons are accelerated on a very short time scale~$t_{\rm
acc} \simeq 10^{-2} \tau_{\rm A}$ (cf.~\cite{Schopper98}).  For this reason
there is no need to consider the precise dynamic evolution of the MHD fields.
We can use them as static background fields without any temporal changes and
this is also a very good approximation.  In addition to the Lorentz
force~$\VV{F}_{\rm L} = q \left( \VV{E} + \VV{v}/c \times \VV{B} \right)$ our
code considers the complete loss force \cite{Landau2}
\begin{eqnarray}
  \VV{F}_{\rm Rad} &=& \frac{2 q^3 \gamma}{3 m c^3} \left\{ \!\! \Dif{\VV{E}}{t} +
     \frac{1}{c}\,\VV{v} \times \Dif{\VV{B}}{t} \right\} \nonumber\\
  &+& \frac{2 q^4}{3 m^2 c^4} \left\{ \frac{1}{q} \VV{F}_{\rm L} \times \VV{B} +
               \frac{1}{c} \, \VV{E} \left( \VV{v} \cdot \VV{E} \right) \right\}
               \nonumber\\
  &-& \frac{2 q^4 \gamma^2}{3 m^2 c^5} \, \VV{v} \left\{ \frac{1}{q^2} F_{\rm
               L}^2 - \frac{1}{c^2} \left( \VV{E} \cdot \VV{v} \right)^2 \right\}, 
\label{equ:loss}
\end{eqnarray}
where $q$, $m$, $\VV{v}$ and $\gamma$ denote the charge, the mass, the velocity
and the Lorentz factor of a particle, respectively.  This loss force describes
all particle-field interactions, inclusive synchrotron radiation (SY) and
inverse Compton scattering (IC).  SY presents the radiation of relativistic
electrons, gyrating around the magnetic field, i.e.~it influences the particle
momentum perpendicular to the magnetic field.  If an acceleration process
injects particles merely along the field lines these particles undergo no
synchrotron losses.  However, since we inject particles isotropically from all
directions, the particles will have finite pitch angles at the beginning.  Even
for very anisotropic energy distributions in the sense that the particles move
only parallel to the magnetic field lines particles may suffer radiative losses
caused by inverse Compton scattering, if an intense photon bath is present in
the environment of the acceleration zone.  The radiation mechanisms depend on
the energy of the particles and the energy density of the magnetic field $U_{\rm
Mag}\propto B^2$ (SY) or the photons $U_{\rm Rad}$ (IC), respectively,

Qualitatively, it is clear that relatively strong (in terms of efficiency and
short time scales) particle acceleration via magnetic reconnection requires
significant magnetic field strengths.  Since astrophysical plasmas are highly
conducting media, strong magnetic fields are related with either fast plasma
motions or high plasma pressure and temperature, i.e intense radiation fields.
Such conditions are typical, e.g., for compact astrophysical objects, as neutron
stars or black holes and magnetospheres of young stellar objects.  Often the
magnetic field energy density and the radiation energy density are of comparable
order, or at least proportional to each other.  This quality is again connected
with the high electric conductivities of astrophysical plasmas on macroscopic
spatial scales.  If plasma is accelerated due to some radiation pressure, the
magnetic field is strongly coupled to the plasma motion and is amplified by
field line compression and twisting.  The plasma temperature is directly related
to the radiation temperature.  Thus, a significant variation of the emitted
radiation will consequently produce varying plasma temperatures and pressures,
which finally leads to magnetic field amplification (e.g.~\cite{Blandford90}).
Thus, the energy losses of the accelerated particles can be handled by a simple
scaling $U_{\rm Mag}\propto U_{\rm rad}$.

\section{Comparison}\label{sec:Comp}

We use our code for the calculation of electron energy spectra which one would
expect to be the final energy distributions in some selected configuration.  In
order to perform this we calculate the trajectories of more than 8000~electrons,
which start at random places with a random momentum and determine their final
energies, when they leave the numerical bounding box of the MHD fields.  The
initial positions are equally distributed in the whole reconnection region.  The
initial momentum of the electrons is chosen isotropic with absolute values that
correspond to a Maxwellian distribution of the initial energy.

\vfill
  \begin{figure}[b]
     \center
     \includegraphics[width=0.8\linewidth,keepaspectratio]{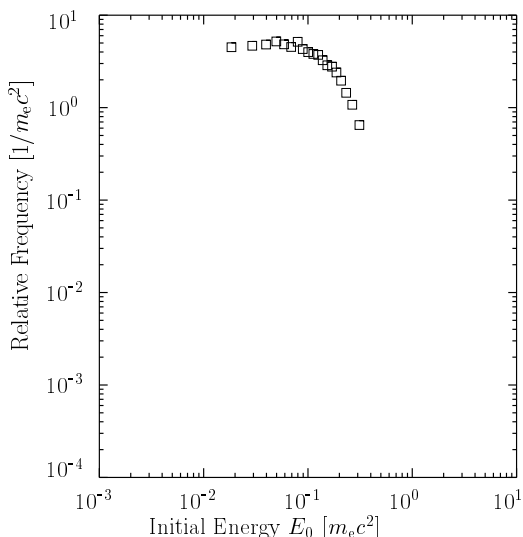}
     \caption{Initial energy distribution (common to all simulations).  }
     \label{fig:inispec}
  \end{figure}
  \begin{figure}[t]
     \center
     \includegraphics[width=0.8\linewidth,keepaspectratio]{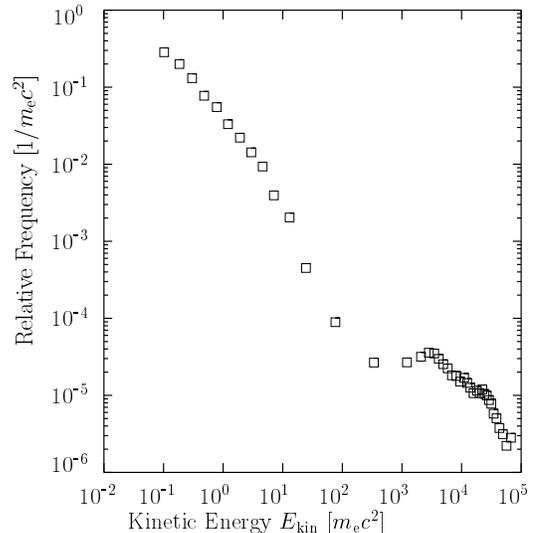}
     \caption{Final energy distribution, calculated with realistic
              3-D-MHD fields.  }
     \label{fig:realistic}
  \end{figure}
In Fig.~\ref{fig:inispec} the initial energy distribution of the electrons is
shown.  Here and in all following plots "Relative Frequency" is the
normed~($\int dE = 1$) percentage of particles per energy interval.  All runs
start from this distribution.  Fig.~\ref{fig:realistic} shows the final energy
distribution of electrons accelerated in our quite realistic 3D-MHD fields (see
Sec.~\ref{sec:MHD}).  Most electrons are accelerated due to the strong guiding
component~$B_z$.  Energies in the range from~$10^{-1} m_{\rm e}c^2$ to
nearly~$10^5 m_{\rm e}c^2$ are seen.  One first remarkable fact is the existence
of electrons nearly without any energy gain.  A second interesting feature is
the obvious depression in the middle of this spectrum.  
  \begin{figure}[b]
     \center
     \includegraphics[width=0.8\linewidth,keepaspectratio]{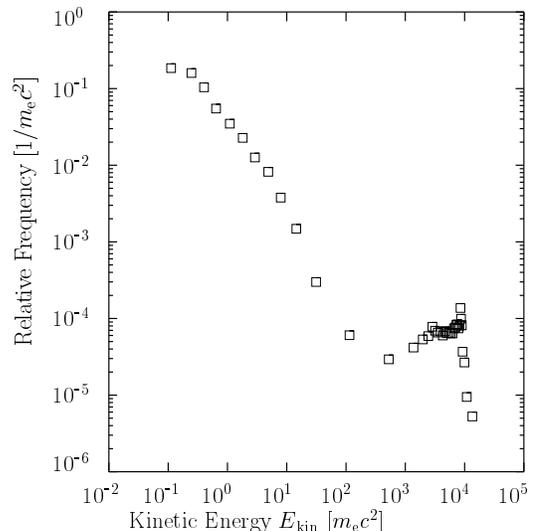}
     \caption{Final energy distribution, calculated with realistic
              3-D-MHD fields and consideration of loss processes.
              $U_{\rm Rad} \sim 10 \cdot U_{\rm Mag}$.  }
     \label{fig:loss}
  \end{figure}
  \begin{figure}[t]
     \center
     \includegraphics[width=0.8\linewidth,keepaspectratio]{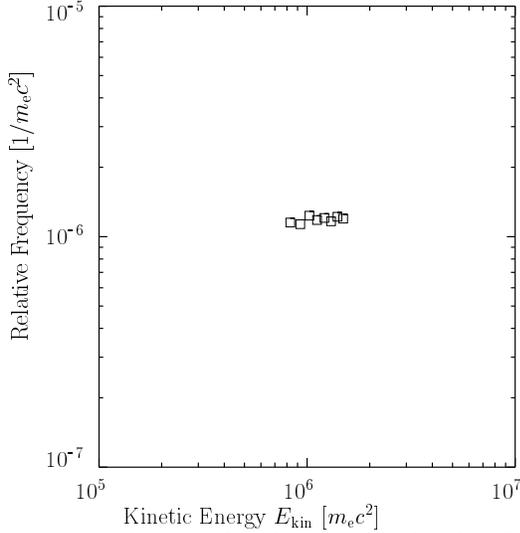}
     \caption{Final energy distribution, calculated with a simplified,
              homogeneous electric field and the realistic magnetic
              field.  }
     \label{fig:ehom}
  \end{figure}
It seems as if there exists some sort of selection process, which does not allow
a particle to achieve an energy between~$10^2 m_{\rm e}c^2$ and~$10^3 m_{\rm
e}c^2$.  This might indicate that there are unstable trajectories, which lead
exactly to this energy range.  The high energetic end of the particle
distribution is determined by the thickness of the resistive region.  Only in
the region of finite resistivity a parallel electric field is present, which can
accelerate the particles efficiently.  
\vfill
  \begin{figure}[b]
     \center
     \includegraphics[width=0.8\linewidth,keepaspectratio]{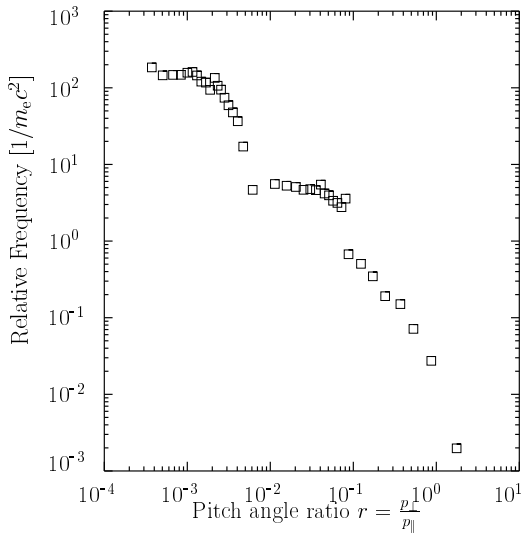}
     \caption{Pitch angle distribution of the electrons simulated in
              the realistic 3D-MHD fields.  }
     \label{fig:pitch_r}
  \end{figure}
  \begin{figure}[t]
     \center
     \includegraphics[width=0.8\linewidth,keepaspectratio]{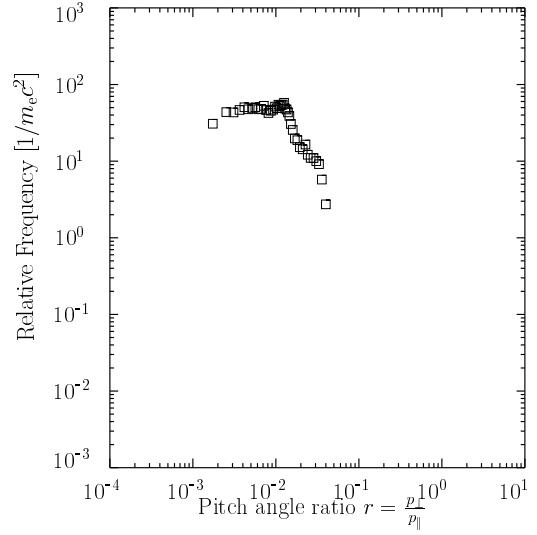}
     \caption{Pitch angle distribution of the electrons simulated in
              the oversimplified electric field.  }
     \label{fig:pitch_x}
  \end{figure}
The gyro radius of an accelerated electron is given by
\begin{equation}
   r_{\rm ce} = v_{\rm Te}\omega_{\rm ce} = \frac{v_{\rm Te}\gamma
   m_{\rm e}c}{eB},
\end{equation}
which depends on the energy of the electrons ($v_{\rm Te}$ is the thermal
velocity of the electrons).  Particles are accelerated, in the best case, as
long as their gyro radii are smaller than the extent of the resistive region.
This criterion gives the maximum energy gain achievable.  The effect of
considering the loss force (Equ.~\ref{equ:loss}) is shown in
Fig.~\ref{fig:loss}.  In order to illustrate those effects we overestimate the
strength of the loss force by assuming an extremely 
  \begin{figure}[b]
     \center
     \includegraphics[width=0.8\linewidth,keepaspectratio]{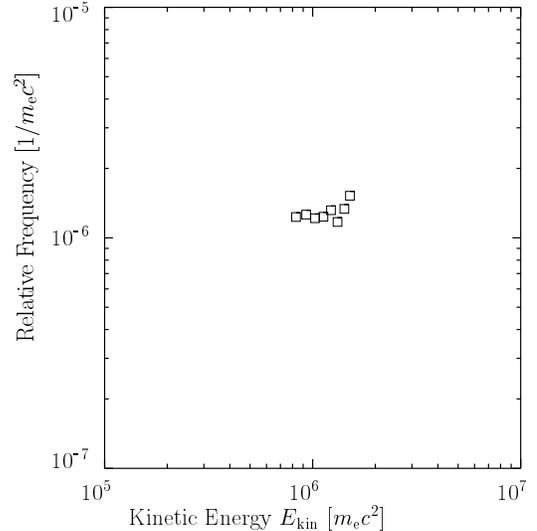}
     \caption{Final energy distribution, calculated with both, an
              homogeneous electric field and a homogeneous magnetic
              field.  }
     \label{fig:ehombhom}
  \end{figure}
intense photon bath with an
energy density~$U_{\rm Rad}$ of about ten times the maximum magnetic energy
density~$U_{\rm Mag}$.  Very few electrons are now able to achieve more
than~$10^{4} m_{\rm e} c^2$.  This is the energy, where the loss of energy due
to radiative processes balances the energy gain from the electric field.  The
characteristic features mentioned before remain unchanged apart from the new
upper limit for the final energy, which leads to a "pile-up" at about~$10^{4}
m_{\rm e} c^2$.

After all, it can be stated that the usage of rather realistic 3D-MHD
fields, i.e. a field configuration that represent evolved reconnection
electromagnetic fields rather than analytical idealized fields,
gives a rich spectrum displaying a great variety of features.  This changes
drastically in the case of an oversimplified homogeneous electric field.  The
next example is calculated with the electric field
\begin{equation}
   \VV{E} \left( \VV{r} \right) =
   \left(
      \begin{array}{c}
         0 \\
         0 \\
         E_{z,\rm max}
      \end{array}
   \right),
\end{equation}
where~$E_{z,\rm max}$ denotes the maximum of the $z$~component of the numerical
electric field used before.  The resulting energy distribution, calculated in
the same way as before, is shown in Fig.~\ref{fig:ehom}.  We should point out
that for this simulation nothing has been changed in the magnetic field
configuration.  Fig.~\ref{fig:ehom} shows a very narrow spectrum around the
energy~$10^6 m_{\rm e}c^2$ completely without \emph{any} feature so
characteristic for the spectrum of Fig.~\ref{fig:realistic}.  First of all, we
can clearly see that every electron is accelerated.  This unphysical result
comes at no surprise, because a constant electric field~$E_z$ in combination
with the strong guiding component~$B_z$ implies an infinite resistive region.
The obvious absence of any structure shows, that the origin of the depression
has to lay mainly in the electric field.  An oversimplification of~$\VV{E}$
neglects this interesting effect.  The enlargement of the resistive region to
the whole numerical box also erases the upper limit of acceleration.  The
maximum energy is given by the potential difference between on side of the box
and the other, but that is a pure trivial geometric limit.

In addition, there is no similarity between the rather realistic spectrum
Fig.~\ref{fig:realistic} and that calculated with a homogeneous electric field
Fig.~\ref{fig:ehom}.

Comparison of the pitch angle distribution~(Fig.~\ref{fig:pitch_r}) of the
realistic simulation with the one of the oversimplified electric field
simulation~(Fig.~\ref{fig:pitch_x}) also shows a striking difference.  While
both distributions clearly show that a vast majority of electrons is moving
mainly along the magnetic field lines, again a broad spectrum over tree decades
and a small amount of electrons, which have a pitch angle greater than
$\frac{\pi}{2}$ evolves (Fig.~\ref{fig:pitch_r}).  The pitch angle distribution
in Fig.~\ref{fig:pitch_x} on the other side is lacking any hint of the electron
population with an pitch angle greater than $\approx 0.01$.  An homogeneous
electric field can, in fact, not be regarded as a valid approximation of a
realistic field configuration.

At this point the question arises, how important is the structure of the
magnetic field for the final spectra anyway?  In order to answer this question
we finally consider the spectrum for
\begin{equation}
   \VV{E} \left( \VV{r} \right) =
   \left(
      \begin{array}{c}
         0 \\
         0 \\
         E_{z,\rm max}
      \end{array}
   \right)\quad
   \VV{B} \left( \VV{r} \right) =
   \left(
      \begin{array}{c}
         0 \\
         0 \\
         B_{z,\rm max}
      \end{array}
   \right)
\end{equation}

This spectrum (Fig.~\ref{fig:ehombhom}) shows a striking similarity with the
previously shown one Fig.~\ref{fig:ehom}.  Both spectra are dominated by the
geometry of the numerical box.  All of the features seen
in~Fig.~\ref{fig:realistic} are coming from the complicated structure of the
electric field.  It seems that the structure of the magnetic field serves only
as "stage" for the electrons without significant influence on the spectra.
Nevertheless, we should keep in mind, that the $\VV{B}$~field determines the
structure of the $\VV{E}$~field.

\section{Discussion}\label{sec:Disc}

Astrophysical plasmas like the interstellar or intergalactic medium and compact
systems like accreting stellar objects are magnetized and highly conducting
collisionless media under the influence of external forces.  The main driving
forces for plasma-magnetic field motions are gravity, rotation and pressure
gradients.  Due to the lack of significant particle collisions astrophysical
plasmas are easily forced to steep magnetic field gradients, which ultimately
will evolve into relaxed field configurations via magnetic reconnection.  In that
sense magnetic reconnection is an unavoidable result of external distortions in
collisionless magnetized plasmas.  The important observable consequences of
reconnection for astrophysical applications follows from the fact that it
accounts for heating and particle acceleration by locally converting the stored
magnetic energy into bulk heating of the plasma via Ohmic heating or
acceleration of a population of high energy particles via magnetic field-aligned
electric fields that are associated with the finite conductivity.

The heating capabilities of magnetic reconnection have been
investigated in detail, since
the heating rate $\propto \eta j^2$ is a macroscopic quantity
which depends only on the density and temperature, but not on the details
of the single particle response to the electromagnetic configuration
inside a reconnection zone.  In that sense it behaves like a macroscopic
transport property.

The acceleration of particles instead depends on the local properties of the
reconnection region.  The energization of a charged particle above all depends on
the effective electric force the particle "feels".  Depending on its initial
position and initial energy the particle energy gain differs.  Since the only
source of information of high energy astrophysics is the electromagnetic
radiation, the energy distribution of relativistic particles is one of the most
required issues concerning the interpretation of nonthermal radiation
sources.  Thus, for any acceleration process discussed it is most important to
investigate the resulting energy distribution functions in dependence on the
various parameters on which the acceleration process depends on.  In this
context, the two-dimensional models published so far for the acceleration of
charged particles by magnetic reconnection are not satisfying, since they
describe acceleration in linear conductors.  They do not describe the effect of
the finite extension of the acceleration zone and show always a purely
monoenergetic distribution at the maximum energy the particles are "allowed" to
reach.  The maximum energy in these models is simply given by the product of the
homogeneous electric field times the chosen length of the reconnection zone,
which is supposed to be a free parameter.  This is definitely incorrect; the
system size must be finite and must depend on the process responsible for the
violation of ideal conductivity.

We have shown the pitfalls of the two-dimensional ansatz and present
3D-simulations of MHD-reconnection plus test particle simulations within the
MHD-frame, including radiation losses (synchrotron and inverse Compton
scattering).  The precise structure of the electric field is the determining
factor of the acceleration process in the MHD reconnection region
(cf.~Sec.~\ref{sec:Comp}).  One has to be extremely careful in simplifying the
electric field; in our example the spectrum calculated with a homogeneous
electric field is no good approximation to the one that results from a
physically motivated field.

Since the investigated magnetic field configuration is a valid 3D generalization
of a 2D X-type field the usage of an homogeneous electric fields in X-type
electric fields should also be misleading as far as the spectrum is concerned.

Whereas particle acceleration in a simplified X-type configuration with a
homogeneous electric field is an interesting subject of it's own, it is
obviously not appropriate for the study of particle acceleration in magnetic
reconnection zones.  We have shown that for this task we need much more
realistic fields and at least numerical approaches in future should use
3D-MHD fields.  A further extension of the work presented in this
contribution will be the consideration of multiple X-line and O-point
reconnection configurations (c.f.~\cite{Kliem98} and \cite{Schumacher96} for
the 2D case) which should show a larger class of different kinds of particles
populations.

The strong influence of radiative losses on the resulting energy distribution of
the accelerated charged particles, even without overestimation of the photon
density, gives us the clear indication that we should not restrict ourselves to
the use of realistic field configurations alone.  By not taking into account SY
and/or IC losses in relativistic particle acceleration processes in reconnection
zones the particle pile-up at the upper energy limit is completely neglected and
the maximum energy of the particles is incorrectly estimated.  This shows that
the consideration of loss processes should be obligatory for the determination
of quantitative results in the relativistic case.  In this context we briefly
note that particle acceleration by 3D-reconnection has considerable
observational consequences. As is obvious from our simulations 3D-reconnection
produces power-law energy distributions, which would emit power-law radiation
spectra, i.e.  decreasing radiation flux with increasing frequency~$\nu$:
Flux~$\propto \nu^{-\alpha}$.  For synchrotron and inverse Compton emission the
index of the energy distribution $p$ is related with the spectral index $\alpha$
via $p=2\alpha +1$ In contrast the simple analytical models with a constant
electric field give mononergetic energy distributions which give synchrotron
spectra where the received radiation flux increases with increasing frequency up
to some critical frequency determined by the energy of the accelerated particles
(Flux~$\propto\nu^{1/3}$ for an optically thin monoenergetic source)
(c.f.~\cite{Longair81}).

At the present time we do not consider any interaction between particles and
plasma waves and we neglect losses due to pair production.  These effect should
be addressed in future simulation studies as well as the interplay of second
order magnetic fields induced by accelerated particle beams and the
ponderomotive force of the excited plasma fluctuations.  We feel that such
studies will prove to be very helpful to clarify the role of reconnection in a
variety of dynamical astrophysical plasma phenomena as different as, e.g.,
extragalactic jets, TeV burst and solar flares.

\section*{Acknowledgments}

This work was supported by the Deutscher Akademischer Austausch Dienst~(DAAD)
through the grant D/98/02999, the Alexander von Humboldt foundation through the
Feodor Lynen program (GTB) and the Deutsche Forschungsgemeinschaft through the
Schwerpunkt "Physik der Sternentstehung" and the grant Le 1939-4/1.

\end{document}